\documentclass[sigconf,nonacm]{acmart}

\setcopyright{acmlicensed}
\copyrightyear{2024}
\acmYear{2024}
\acmDOI{XXXXXXX.XXXXXXX}

\acmConference[Conference acronym 'XX]{Make sure to enter the correct
  conference title from your rights confirmation email}{June 03--05,
  2018}{Woodstock, NY}
\acmISBN{978-1-4503-XXXX-X/18/06}

\usepackage{listings}
\usepackage{tikz}
\usepackage{pgfplots}
\pgfplotsset{compat=1.18} 
\usetikzlibrary{matrix,fadings,calc,positioning,decorations.pathreplacing,decorations.text,arrows,patterns,plotmarks,babel,mindmap,trees,shadows,tikzmark}
\usepgfplotslibrary{fillbetween}
\newcommand{\tikzextchoice}[2]{#1} 
\tikzextchoice{}{\tikzexternalize}

\usepackage{siunitx}
\usepackage{multirow}
\usepackage{xcolor}

\newenvironment{customlegend}[1][]{%
    \begingroup
    \csname pgfplots@init@cleared@structures\endcsname
    \pgfplotsset{#1}%
}{%
    \csname pgfplots@createlegend\endcsname
    \endgroup
}%
\def\addlegendimage{\csname pgfplots@addlegendimage\endcsname}

\newcommand{\addlegendimageintext}[1]{%
    \tikz {
        \begin{customlegend}[
            legend entries={\empty},
            legend style={
                draw=none,
                inner sep=0pt,
                column sep=0pt,
                nodes={inner sep=0pt}}]
        \addlegendimage{#1}
        \end{customlegend}
    }%
}

\begin{document}

\newcommand{\RR}{{\mathbb R}}
\newcommand{\EE}{{\mathbb E}}
\newcommand{\PP}{{\mathbb P}}
\newcommand{\dd}{{\mathrm d}}
\newcommand{\Var}{{\text{Var}}}
\newcommand{\OO}{{\mathcal O}}
\newcommand{\dotequal}{:=}
\newcommand{\expected}[1]{\EE #1}
\newcommand{\samplemean}[1]{\overline{#1}}
\newcommand{\expectedvaluediff}{\expected{(f')}}
\newcommand{\averagediff}{\samplemean{f'}}
\newcommand{\diffexpectedvalue}{(\expected{f})'}

\title{Optimization Using Pathwise Algorithmic Derivatives of Electromagnetic Shower Simulations}

\author{Max Aehle}
\email{max.aehle@scicomp.uni-kl.de}
\orcid{0000-0002-6739-5890}
\affiliation{%
  \institution{University of Kaiserslautern-Landau (RPTU)}
  \streetaddress{Gottlieb-Daimler-Straße}
  \city{Kaiserslautern}
  \country{Germany}
}

\author{Mihály Novák}
\email{mihaly.novak@cern.ch}
\affiliation{%
  \institution{European Organization for Nuclear Research}
  \country{(CERN)}
}

\author{Vassil Vassilev}
\email{vassil.vassilev@cern.ch}
\affiliation{%
  \institution{Princeton University}
  \city{Princeton}
  \state{New Jersey}
  \country{USA}
}

\author{Nicolas R.\ Gauger}
\email{nicolas.gauger@scicomp.uni-kl.de}
\orcid{0000-0002-5863-7384}
\affiliation{%
  \institution{University of Kaiserslautern-Landau (RPTU)}
  \city{Kaiserslautern}
  \country{Germany}
}

\author{Lukas Heinrich}
\email{l.heinrich@tum.de}
\affiliation{%
  \institution{TU Munich}
  \city{Munich}
  \country{Germany}
}

\author{Michael Kagan}
\email{makagan@slac.stanford.edu}
\affiliation{%
  \institution{SLAC National Accelerator Laboratory}
  \city{Menlo Park}
  \state{CA}
  \country{USA}
}

\author{David Lange}
\email{david.lange@cern.ch}
\affiliation{%
  \institution{Princeton University}
  \city{Princeton}
  \state{New Jersey}
  \country{USA}
}

\renewcommand{\shortauthors}{Aehle et al.}

\begin{abstract}
  Among the well-known methods to approximate derivatives of expectancies computed by Monte-Carlo simulations, averages of pathwise derivatives are often the easiest one to apply. Computing them via algorithmic differentiation typically does not require major manual analysis and rewriting of the code, even for very complex programs like simulations of particle-detector interactions in high-energy physics. However, the pathwise derivative estimator can be biased if there are discontinuities in the program, which may diminish its value for applications. 
  
  This work integrates algorithmic differentiation into the electromagnetic shower simulation code HepEmShow based on G4HepEm, allowing us to study how well pathwise derivatives approximate derivatives of energy depositions in a sampling calorimeter with respect to parameters of the beam and geometry. %
  We found that when multiple scattering is disabled in the simulation, means of pathwise derivatives converge quickly to their expected values, and these are close to the actual derivatives of the energy deposition. Additionally, we demonstrate the applicability of this novel gradient estimator for stochastic gradient-based optimization in a model example.
\end{abstract}


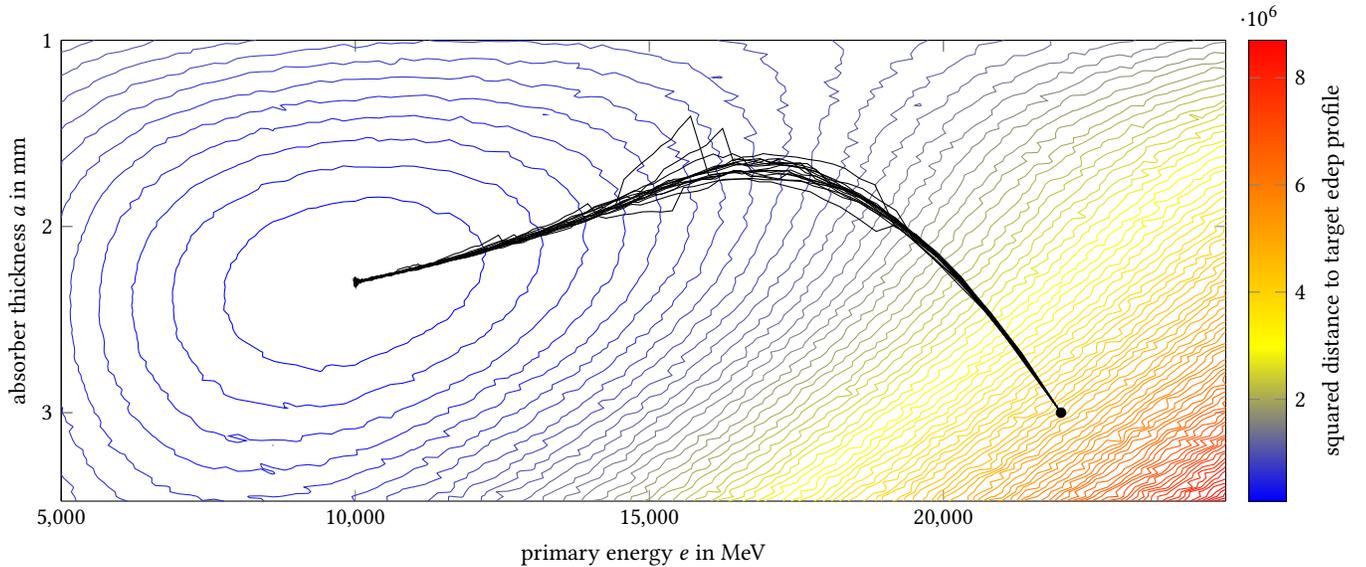
\begin{teaserfigure}
 \tikzextchoice{\includegraphics{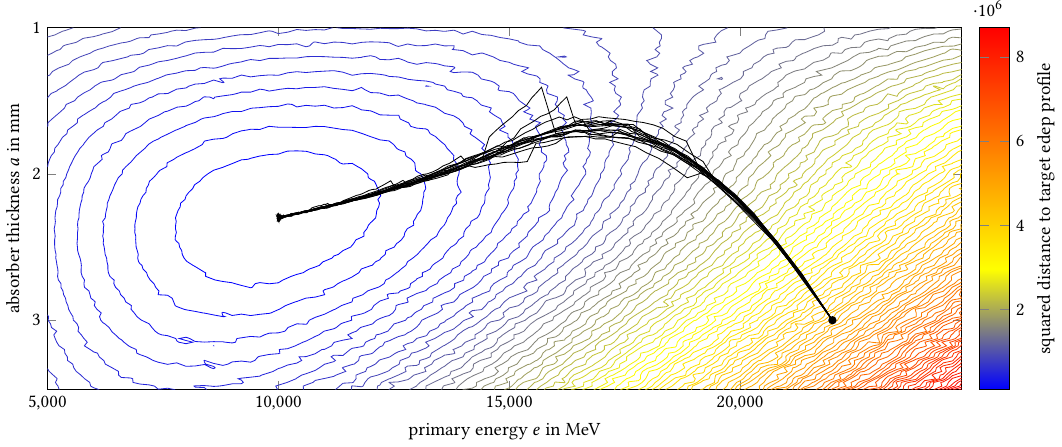}}{
    \tikzsetnextfilename{optimization}
\begin{tikzpicture}
\begin{axis}[ylabel={primary energy $e$ in \si{\mega\eV}},ylabel style={align=center},xlabel={absorber thickness $a$ in \si{\milli\meter}},height=0.3446\textwidth,width=0.87\textwidth,view={90}{90},colorbar,colorbar style={ylabel={squared distance to target edep profile},ylabel near ticks, yticklabel pos=right}, scale only axis=true, scaled ticks=false, tick label style={/pgf/number format/fixed}, xtick={1.0,2.0,3.0}, ytick={5000,10000,15000,20000,25000},xlabel style={rotate=90}  ]
\addplot3[mesh/rows=100,mesh/num points=10000,contour gnuplot={
                    number=100,
                    output point meta=rawz,
                    labels=false,
                }] table [x index={0}, y index={1}, z index={2}] {images/landscape-ea.dat};
\fill[black] (axis cs:3,22000) circle(2pt);
\foreach\i in {0,1,...,15}{
  \addplot[black,thin] table[x index=2, y index=1] {images/convhist/convhist-\i.txt};
}
\end{axis}
\end{tikzpicture}
}
  \caption{Reconstruction of the values of primary energy and absorber thickness that lead to a given energy deposition profile in a sampling calorimeter, using the gradient descent optimizer with algorithmic derivatives of the shower simulation.}
  \label{fig:optimization}
\end{teaserfigure}


\maketitle

\section{Introduction}

{\bfseries Monte-Carlo simulations.} Monte-Carlo (MC) simulations are a popular method to model processes that involve stochasticity; for instance, the Geant4 toolkit  \cite{agostinelli_geant4simulation_2003,allison_geant4_2006,allison_recent_2016} is widely used to simulate the passage of particles through matter. Unlike deterministic simulations, the output data $y \in Y \subset \RR^m$ of MC simulations does not only depend on the input data $\theta\in \Theta \subset \RR^n$, but also on random numbers supplied by a pseudo-random number generator (RNG). We can think of MC simulations as functions
\begin{equation}\label{eq:mc-as-function}
    f: \Theta \times \Omega \to Y, (\theta, \omega) \mapsto y
\end{equation}
with an additional argument $\omega$ from a probability space $\Omega$ with a probability measure $\PP$. For simplicity, we assume in the following that the RNG defines only a single stochastic primitive called {\ttfamily flat()} (as in Geant4) that returns independent random numbers uniformly distributed on the interval $[0,1]$, like {\ttfamily numpy.random.rand} in Python or {\ttfamily (double)rand()/RAND\_MAX} in C. We may think of $\Omega$ as the set of sequences of random numbers.

Usually, the function $f$ is evaluated many times; a common quantity of interest for a MC simulation is the expected value of the output for a given input $\theta$,
\begin{equation}\label{eq:expected-value} 
\expected{f} \dotequal \EE_\omega f(\theta,\omega) = \int f(\theta, \omega) \,\mathrm d \PP(\omega),
\end{equation} 
which can be estimated by averaging over $N$ independent random samples,
\begin{equation}\label{eq:average}
  \samplemean{f} \dotequal  \frac{1}{N} \cdot \sum_{i=1}^{N} f(\theta,\omega^{(i)}).
\end{equation}
The choice of $N$ must balance the required run-time, which grows linearly with $N$, with the standard deviation of $\samplemean{f}$, which is proportional to $N^{-1/2}$.

{\bfseries Algorithmic Differentiation.} Sometimes, users of (for now, deterministic) computer simulations are not primarily interested in the output $y$ at a specific input value $\theta$, but rather wish to identify \emph{optimal inputs} $\theta \in \Theta \subset \RR^n$ that maximize or minimize a scalar output $y$. For example, $\theta$ might be a set of parameters to be tuned in order to minimize the deviation $y$ between model predictions and observed data; closely related, $\theta$ might contain the weights of a neutral network and $y$ be the training error. As another example, $\theta$ might be a set of design parameters and $y$ a utility function to be improved, see e.\,g.\ the work of Albring et al. \cite{doi:10.2514/6.2016-3518} who optimized the shape of an airfoil to reduce the drag computed by a computational fluid dynamics simulation. 

To employ \emph{gradient-based optimization} methods like gradient descent or BFGS \cite{shanno_conditioning_1970}, it is necessary to be able to evaluate gradients $\partial y/\partial \theta$. Besides, the derivative $\partial y / \partial \theta$ characterizes the sensitivity of $y$ with respect to changes in $\theta$, and can thus be useful for uncertainty quantification. When the function $\theta\mapsto y$ is given by computer code, the gradient $\partial y/\partial \theta$ of a computer-implemented deterministic function can often be obtained efficiently and accurately by \emph{algorithmic differentiation (AD)} \cite{griewank_evaluating_2008}, a set of techniques based on the chain rule and the well-known derivatives of the elementary operations performed by the  computer program while evaluating $\theta \mapsto y$. 

Specifically, the \emph{forward mode} of AD with a single scalar input $\theta \in \RR$ (i.\,e.\ $n=1$) keeps track of the \emph{dot value} $\dot q = \partial q / \partial \theta$ whenever an intermediate variable $q$ is computed by the program; for example, the \emph{primal} operation $q = q_1 \cdot q_2$ is augmented with the \emph{AD logic} $\dot q = \dot q_1\cdot q_2 + q_1 \cdot \dot q_2$. Optimization applications favor the \emph{reverse mode} of AD because it provides the entire gradient of a scalar output $y \in \RR$ (i.\,e.\ $m=1$) with respect to many inputs $\theta\in\RR^n$ in a run-time independent from $n$, however at the expense of a higher memory consumption; we refer to the textbook by Griewank and Walther \cite{griewank_evaluating_2008} for details. On the implementation side, there are several  mechanisms for \emph{AD tools} to detect real arithmetic in an existing primal program and to augment them with AD logic; for instance, \emph{operator-overloading tools} provide a custom floating-point datatype with arithmetic operators and math functions overloads, to be used instead of the built-in floating-point types like {\ttfamily double} in C{\ttfamily++}. %
In contrast, \emph{source transformation} tools operate on the program as a whole. While they are usually more difficult to implement and may only support a subset of the language, having access to the entire program allows for more advanced performance optimizations.

{\bfseries AD for MC Simulations.} Typically, AD tools recognize and differentiate the basic operations $+$, $-$, $\cdot$ and $/$ and related operators like {\ttfamily +=}, as well as simple math functions like $\sqrt{\phantom{a}}$, $\exp$, $\sin$, etc. Higher-level mathematical constructs often need manual treatment; while it is usually straightforward to inform AD with analytical derivatives of, e.\,g., solutions of linear systems \cite{FISCHER1991169} and the dominating eigenvalue of a  matrix \cite{PhysRevB.101.245139}, computing the derivative of an expected value of a MC simulation,
\begin{equation}\label{eq:diff-expected-value} 
\diffexpectedvalue \dotequal \frac{\partial}{\partial \theta} \left[ \EE_\omega f(\theta,\omega) \right] = \frac{\partial}{\partial \theta} \left[ \int f(\theta,\omega)\,\mathrm d \PP(\omega) \right],
\end{equation}
poses a rather difficult but very important challenge across application domains. 

In quantitative finance,  certain derivatives of e.\,g.\ expected option prices are called ``Greeks'' and define strategies to hedge risks \cite{Glasserman2003Chapter7}. Differentiable rendering allows to reconstruct three-dimensional scenes from images \cite{Li:2018:DMC,bangaru2020warpedsampling,Zeltner2021MonteCarlo,BangaruMichel2021DiscontinuousAutodiff,10.1145/3649843}. In reinforcement learning, policy gradients can be used for training \cite{10.5555/3009657.3009806}. In many of the aforementioned application areas of gradient-based optimization using deterministic AD, it is natural to add stochasticity to the differentiated code, leading to e.\,g.\ stochastic neural networks \cite{NIPS2013_d81f9c1b} including VAEs \cite{DBLP:journals/corr/KingmaW13} and GANs \cite{NIPS2014_5ca3e9b1}. See \cite{mohamed2020monte} for a review of Monte Carlo gradient estimation in machine learning.

The present work is a study on applying AD in the realm of high-energy physics (HEP), where gradient-based optimization is explored as a way to enhance the design of future particle detectors \cite{DORIGO2023100085,aehle2023progress} or reconstruct properties of detected particles \cite{DECASTRO2019170,Simpson_2023}, and gradients of stochastic programs could help performing Bayesian inference of parameters of the standard model \cite{10.1145/3295500.3356180}. 
The Geant4 toolkit for the simulation of the passage of particles through matter \cite{agostinelli_geant4simulation_2003,allison_geant4_2006,allison_recent_2016} is widely used across many HEP-related application areas, from the planning of detectors at the LHC to radiation safety in space to medical physics.

As a step to explore ways to create a differentiated version of Geant4, in this study, we differentiate a more compact but algorithmically similar MC code composed of G4HepEm \cite{g4hepem-github} and HepEmShow \cite{hepemshow-github,hepemshow-doc}. We are interested in derivatives of the expected value of energy deposition of electromagnetic showers in a simple sampling calorimeter, with respect to parameters of the geometry and the incoming particles.

A natural first step to approach \eqref{eq:diff-expected-value} is to form the \emph{pathwise derivative} 
\begin{equation}\label{eq:pathwise-diff}
\frac{\partial }{\partial \theta} f(\theta,\omega)
\end{equation}
by applying AD to the MC simulation $f$ in a way that, with regard to differentiation, treats random numbers like constants. This has been accomplished for Geant4 in principle, without focus on performance though and only simulating a single particle to demonstrate technical feasibility \cite{aehle2023progress}. The second step then is to estimate the expected value of the pathwise derivative, \begin{equation}\label{eq:expected-value-diff} 
 \expectedvaluediff \dotequal \EE_\omega \left[ \frac{\partial}{\partial \theta} f(\theta,\omega) \right],
\end{equation}
by averaging it over $N_\text{diff}$ independent random samples,
\begin{equation}\label{eq:average-diff}
\averagediff \dotequal \frac{1}{N_\text{diff}} \cdot \sum_{i=1}^{N_\text{diff}} \left[ \frac{\partial}{\partial \theta} f(\theta,\omega^{(i)}) \right].
\end{equation}

\begin{figure}

\begin{minipage}{0.4\linewidth}
\tikzextchoice{\includegraphics{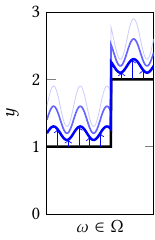}}{
    \tikzsetnextfilename{pathwise_derivative_model_1}
\begin{tikzpicture}
\begin{axis}[xmin=0,xmax=1,ymin=0,ymax=3,domain=0:1, xlabel={$\omega \in \Omega$}, ylabel={$y$}, ylabel near ticks, xlabel near ticks, width=\linewidth, height=5cm,xtick=\empty]
\addplot[black,very thick,samples=1000] {1+(x>0.6)*1};
\addplot[blue!20,samples=1000] {1+(x>0.6)*1+0.6+0.3*sin(1440*x)};
\addplot[blue!60,thick,samples=1000] {1+(x>0.6)*1+0.4+0.2*sin(1440*x)};
\addplot[blue,very thick,samples=1000] {1+(x>0.6)*1+0.2+0.1*sin(1440*x)};
\begin{scope}[blue,->]
\draw (axis cs: 0.1,1.0) -- (axis cs:0.1,1.2587785252292473);
\draw (axis cs: 0.2,1.0) -- (axis cs:0.2,1.10489434837048464);
\draw (axis cs: 0.3,1.0) -- (axis cs:0.3,1.29510565162951535);
\draw (axis cs: 0.4,1.0) -- (axis cs:0.4,1.14122147477075273);
\draw (axis cs: 0.5,1.0) -- (axis cs:0.5,1.2);
\draw (axis cs: 0.7,2.0) -- (axis cs:0.7,2.10489434837048463);
\draw (axis cs: 0.8,2.0) -- (axis cs:0.8,2.29510565162951535);
\draw (axis cs: 0.9,2.0) -- (axis cs:0.9,2.14122147477075275);
\end{scope}
\end{axis}
\end{tikzpicture}
}
\end{minipage}
\begin{minipage}{0.55\linewidth}
\begin{lstlisting}[language=c,basicstyle=\ttfamily]
double f_1(double theta){
  double r = flat();
  double offset = theta * 
    (d+sin(8*pi*r));
  if(r<0.6){
    return 1.0+offset;
  } else {
    return 2.0+offset;
  }
}
\end{lstlisting}
\end{minipage}
\begin{minipage}{0.4\linewidth}
\tikzextchoice{\includegraphics{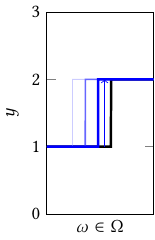}}{
    \tikzsetnextfilename{pathwise_derivative_model_2}
\begin{tikzpicture}
\begin{axis}[xmin=0,xmax=1,ymin=0,ymax=3,domain=0:1, xlabel={$\omega \in \Omega$}, ylabel={$y$}, ylabel near ticks, xlabel near ticks, width=\linewidth, height=5cm,xtick=\empty]
\addplot[black,very thick,samples=1000] {1+(x>0.6)*1};
\addplot[blue!20,samples=1000] {1+(x>0.24)*1};
\addplot[blue!60,thick,samples=1000] {1+(x>0.36)*1};
\addplot[blue,very thick,samples=1000] {1+(x>0.48)*1};
\begin{scope}[blue,->]
\draw (axis cs: 0.54,1.0) -- (axis cs:0.54,2.0);
\end{scope}
\end{axis}
\end{tikzpicture}
}
\end{minipage}
\begin{minipage}{0.55\linewidth}
\begin{lstlisting}[language=c,basicstyle=\ttfamily]
double f_2(double theta){
  double r = flat();
  if(r<0.6-d*theta){
    return 1.0;
  } else {
    return 2.0;
  }
}
\end{lstlisting}
\end{minipage}

\begin{minipage}{0.9\linewidth}
\begin{tabbing}
\raisebox{0.20em}{\addlegendimageintext{black,very thick}} $f(\theta, \omega)$ \qquad\qquad\qquad \=
\raisebox{0.20em}{\addlegendimageintext{blue,very thick}} $f(\theta+\dd \theta, \omega)$ \\
\raisebox{0.20em}{\addlegendimageintext{blue!60,thick}} $f(\theta+2\dd \theta, \omega)$ \>
\raisebox{0.20em}{\addlegendimageintext{blue!20}} $f(\theta+3\dd \theta, \omega)$ \\
\end{tabbing}
\end{minipage}

    \caption{\boldmath Different mechanisms for Monte-Carlo simulations to combine input parameters and randomness. The RNG primitive {\ttfamily flat()} yields independent random numbers uniformly distributed on $[0,1]$; this is how the second argument~$\omega$ in \eqref{eq:mc-as-function} comes in. Around $\theta=0$, both functions have $\EE_\omega f_i(\theta,\omega) = d\cdot \theta$. Top: The pathwise derivative is distributed around the mean $d$. Bottom: The pathwise derivative is zero almost everywhere, and undefined at a single point where $f$ jumps.}
    \label{fig:pathwise-derivative-model}
\end{figure}
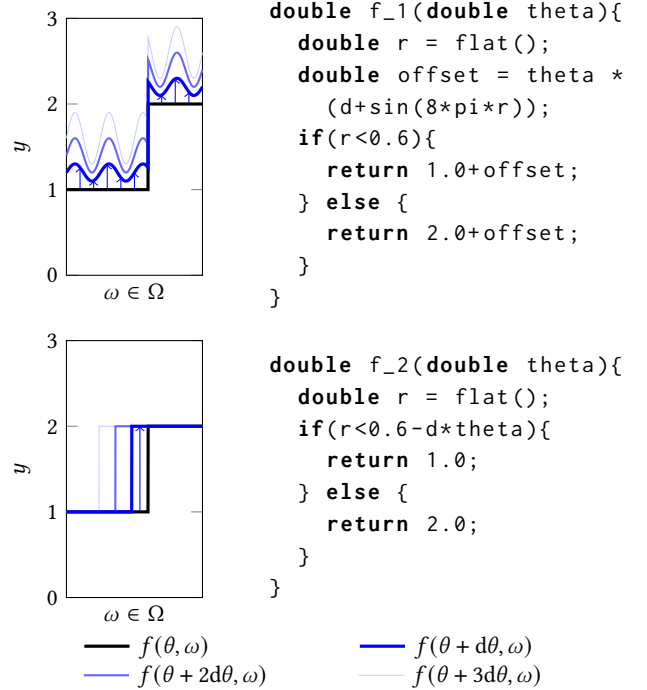

However, the expected pathwise derivative $\expectedvaluediff$ matches the sought derivative $\diffexpectedvalue$ of the expected value only under certain assumptions on $f$. A well-known corollary of Lebesgue's dominated convergence theorem \cite[Theorem~A.5.3]{durrett} states that $\diffexpectedvalue = \expectedvaluediff$ if $f(\theta,\omega)$ is continuously differentiable in $\theta$ and $|\tfrac{\partial f}{\partial \theta}|\leq B(\omega)$ for an integrable random variable $B: \Omega\to \RR$. Figure~\ref{fig:pathwise-derivative-model} gives an example of such a function $f_1$ with $(\EE f_1)'=\EE(f_1')=d$.

In general, $\expectedvaluediff$ and $\diffexpectedvalue$ can take different values. For the function $f_2$ in figure~\ref{fig:pathwise-derivative-model}, we can analytically see that 
\begin{gather*}
    \EE_\omega f_2(\theta,\omega) = (0.6-d\cdot \theta) \cdot 1 + (0.4+d\cdot \theta) \cdot 2 = 1.4 + d \cdot \theta \notag \\
    \Rightarrow \frac{\partial}{\partial \theta}\left[ \EE_\omega f_2(\theta,\omega) \right] = d, 
\end{gather*}
but the pathwise derivative $\tfrac{\partial}{\partial \theta} f_2(\theta,\omega)$ is zero for almost all $\omega$. Only for the zero-probability set of $\omega$ with $r(\omega)=0.6-d\cdot \theta$, $f_2(\theta,\omega)$ has a jump at $\theta$. This jump makes $\diffexpectedvalue$  non-zero but does not affect $\expectedvaluediff$. An estimator like $\samplemean{f'}$, whose expected value does not match the target value $\diffexpectedvalue$, is called \emph{biased}.

Non-trivial MC simulations usually contain control flow constructs like {\ttfamily if} and {\ttfamily while}, whose (discrete and hence non-dif\-fer\-en\-tiable) condition depends on both the AD inputs $\theta$ and the randomness $\omega$, so their pathwise derivative estimators are generally biased. Accordingly, several approaches to create unbiased estimates for derivatives of expected values of MC simulations have been proposed in the literature; see e.\,g.\ references~\cite{arya2022automatic,Lew_2023,BangaruMichel2021DiscontinuousAutodiff}, or Kagan and Heinrich \cite{kagan2023branches} for a first analysis of some of these methods in HEP.

The \emph{reparametrization trick} \cite{DBLP:journals/corr/KingmaW13} refers to implementing parametric random distributions as differentiable expressions of the parameters and non-parametric random numbers. For instance, a random number uniformly distributed on $[0,\theta]$ (with $0<\theta\leq 1$, say) would be implemented as $\theta\cdot\text{\ttfamily flat()}$ rather than, e.\,g., rejection-sampling {\ttfamily flat()} repeatedly until it yields a number in $[0,\theta]$. This makes the differentiable dependency between the sampled random numbers and the parameters visible to the mean pathwise derivative  $\expectedvaluediff$. For elementary parametric distributions like normal distributions, MC simulations in HEP often follow the reparametrization trick by default. However, MC simulations then usually continue to process them using non-differentiable operations (like $f_2$ in figure~\ref{fig:pathwise-derivative-model}), to which the reparametrization trick cannot be applied in general.

As a well-known alternative or addition to pathwise derivatives, the \emph{likelihood ratio} or \emph{score function} method \cite{10.1145/84537.84552,10.5555/2969442.2969633} proposes to compute a term
\begin{equation}\label{eq:score-function}
\EE_\omega\left[ \frac{\partial \log(p(\theta,\omega))}{\partial \theta} \cdot f(\theta,\omega) \right] = %
\EE_\omega\left[ \frac{ \tfrac{\partial p}{\partial \theta}(\theta,\omega)  }{p(\theta,\omega)} \cdot f(\theta,\omega) \right].
\end{equation}
This term accounts for the part of the derivative of $\EE f$ related to a differentiable change of the probability $p(\theta,\omega)$ that discrete random events (e.\,g.\ whether an {\ttfamily if} or {\ttfamily else} branch is taken) turn out in the way they do when $f(\theta,\omega)$ is computed. As such, \eqref{eq:score-function} should be added to $\expectedvaluediff$. 

Indeed, for the function $f_2$ in figure~\ref{fig:pathwise-derivative-model}, the bracketed expression in \eqref{eq:score-function} evaluates to $[\tfrac{-d}{0.6}\cdot 1.0]$ when the {\ttfamily if} branch is taken (\SI{60}{\percent} probability) and to $[\tfrac{d}{0.4}\cdot 2.0]$ when the {\ttfamily else} branch is taken (\SI{40}{\percent} probability), giving an expected value of $d$.
However, we were only able to determine the values of the probabilities $p=0.6, 0.4$ in the denominators, and their derivatives $\tfrac{\partial p}{\partial \theta} = -d, d$ in the numerators, because the condition of the {\ttfamily if} statement is very simple.

In the case of MC particle simulations, it is unclear how to determine $p$, because these simulations typically implement stochasticity by combining several random numbers and variables depending on the AD input $\theta$, in non-linear ways. Additionally, the term \eqref{eq:score-function} tends to have a high variance \cite{pmlr-v32-rezende14}. 

The \emph{stochastic AD} method by Arya et al.\ \cite{arya2022automatic} integrates certain kinds of discrete randomness into pathwise derivatives. For each intermediate value appearing in the MC program, this method keeps track of an alternative value that could have been attained with different random outcomes, and the derivative of the probability of such an outcome with respect to the AD input. While we consider it an interesting and promising approach, it appears not to be easily applicable to a MC particle simulation, because {\ttfamily if} statements and discrete randomness originating from comparisons of continuous random values are not yet supported.

Instead of trying to create an unbiased estimator for $\diffexpectedvalue$, in this work, we analyze the biased estimator $\samplemean{f'}$ for a MC code with full electromagnetic physics coverage but simple geometry.  It turns out that when a single physics process called \emph{multiple scattering} is disabled in our setup, the variance of $\samplemean{f'}$ is sufficiently low to obtain reliable estimates of $\expectedvaluediff$ for moderate $N_\text{diff}$, and $\expectedvaluediff$ deviates from a difference quotient approximation of $\diffexpectedvalue$ only by a few percent. A bias of this magnitude can be perfectly acceptable as the derivatives only serve as a tool to guide optimization algorithms (and are not physical quantities that have to match measurements).

Section~\ref{sec:simulation} gives an overview on the simulated hardware setup and the MC code, which we differentiated following the methodology described in Section~\ref{sec:ad}. We then report on the stochastic noise of the MC code (Section~\ref{sec:noise}), the variance and bias of the pathwise algorithmic derivative estimators (Section~\ref{sec:variance-bias}), and a simple demonstrator using these estimators for gradient-based optimization (Section~\ref{sec:optimization}), closing with conclusions and an outlook in Section~\ref{sec:conclusion}.

\section{Simulation of Electromagnetic Showers in a Sampling Calorimeter}\label{sec:simulation}

\subsection{Detector Geometry}\label{sec:detector-geometry}
\begin{figure}
    \centering
    \includegraphics[width=0.6\linewidth]{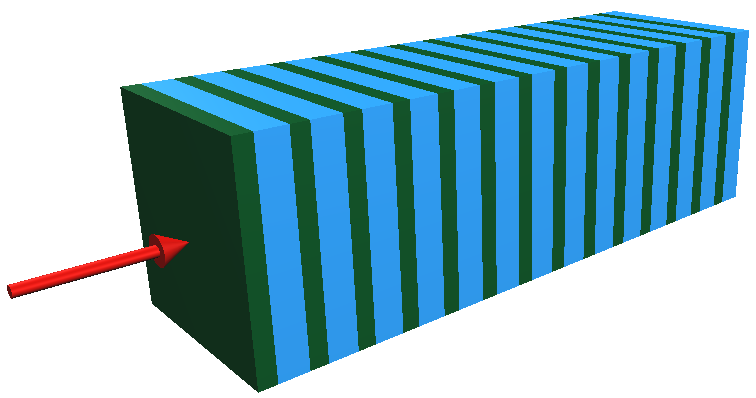}
    \caption{Geometric structure of the sampling calorimeter. Figure courtesy of Nov\'ak et al.\ \cite{hepemshow-github,hepemshow-doc}.}
    \label{fig:geometry}
\end{figure}
Figure~\ref{fig:geometry} shows the simple detector geometry used in this study. The detector hardware is a stack of $n_\text{l}$ identical pairs of \emph{absorber} and \emph{gap layers}, each with a thickness of $a$ and $g$ (respectively) and transversal dimensions $d_t \times d_t$. The two types of layer are each made from homogeneous material; in particular, material properties are piecewise constant and change only at well-known two-dimensional volume boundaries. This assumption on the detector geometry is also made by Geant4 and is usually satisfied in practice. Primary particles arrive centered and orthogonally with an initial kinetic energy $e$.

A default set of values for these parameters is specified in table~\ref{tab:parameters}. The setup was created by Nov\'ak et al.\ \cite{hepemshow-github,hepemshow-doc} and is based on Geant4's TestEm3 test case; however, the absorber material is lead tungstate ($\text{PbWO}_4$) instead of elementary lead (Pb), as a mixture of different atoms makes the test case more general. In this study, the primary energy $e$ and the layer thicknesses $a$ and $g$ will be considered as AD inputs $\theta$, and all other parameters are considered constant.

\begin{table}
    \centering
    \begin{tabular}{lccr}
         \toprule
         Parameter & Symbol  & Arg. & Default value  \\
         
         \midrule
         Kinetic energy of primaries & $e$ & {\ttfamily-e} & \SI{10000}{\mega\eV} \\
         Thickness of absorber layers & $a$ & {\ttfamily-a} & \SI{2.3}{\milli\meter} \\
         Thickness of gap layers & $g$ & {\ttfamily-g} & \SI{5.7}{\milli\meter} \\
         \midrule
         Transversal dimension & $d_t$ & {\ttfamily -t} & \SI{400}{\milli\meter} \\
         Number of layers & $n_\text{l}$ & {\ttfamily-l} & 50 \\
         Type of primary particles & ~ & {\ttfamily-p} & electrons \\
         Absorber material & ~ & {$\tbinom{\text{JSON}}{\text{file}}$} & $\text{PbWO}_4$\\
         Gap material & ~ & {$\tbinom{\text{JSON}}{\text{file}}$} & liquid Ar\\
         \bottomrule
    \end{tabular}
    \caption{Parameters of the simple sampling calorimeter geometry displayed in figure~\ref{fig:geometry}.}
    \label{tab:parameters}
\end{table}

\subsection{Electromagnetic Showers}\label{sec:em-showers}
Electrons, positrons and photons interact with the surrounding matter through various physical processes: ionization, bremsstrahlung, annihilation, pair production, the photoelectric effect, Compton scattering, etc. These processes happen at discrete points in time and can result in a loss of kinetic energy, change of momentum,  deposition of energy in the surrounding matter, and/or creation of \emph{secondary particles}. The interaction rates/\emph{cross-sections} for these processes and their possible outcomes depend on the type and kinetic energy of the particle, and the material composition of the surrounding matter. Except for rare lepto- and photo-nuclear interactions that are neglected in the following, secondary particles are either electrons, positrons and photons. Secondary particles themselves interact with the surrounding matter, forming an electromagnetic shower.

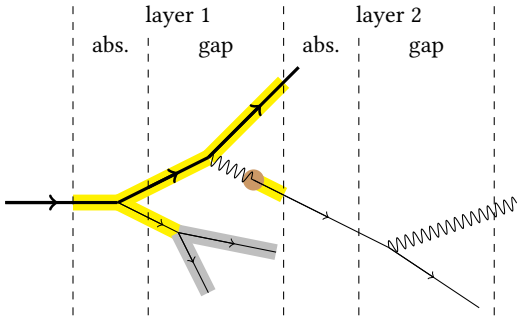
\begin{figure}
    \centering
    \tikzextchoice{\includegraphics{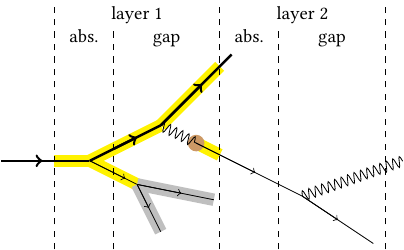}}{
    \tikzsetnextfilename{shower_sketch}
\begin{tikzpicture}
\draw[line width=6pt,yellow] (1.0,1.5) -- (1.6,1.5);
\draw[very thick] (0.1,1.5) -- (1.6,1.5);
\draw[very thick,->] (0.1,1.5)--(0.8,1.5);
\draw[line width=6pt,yellow] (1.6,1.5) -- (2.8,2.1);
\draw[line width=6pt,yellow] (1.6,1.5) -- (2.4,1.1);

\draw (1.6,1.5) -- (2.4,1.1);
\draw[->] (1.6,1.5) -- (2.2,1.2);
\draw[line width=6pt,lightgray] (2.4,1.1) -- (3.7,0.84);
\draw[line width=6pt,lightgray] (2.4,1.1) -- (2.8,0.3);
\draw (2.4,1.1) -- (3.7,0.84);
\draw (2.4,1.1) -- (2.8,0.3);
\draw[->] (2.4,1.1) -- (3.15,0.95);
\draw[->] (2.4,1.1) -- (2.6,0.7);
\draw[line width=6pt,yellow] (2.8,2.1) -- (3.8,3.1);
\draw[very thick] (1.6,1.5) -- (2.8,2.1);
\draw[very thick,->] (2.0,1.7)--(2.4,1.9);
\draw[very thick] (2.8,2.1) -- (4.0,3.3);
\draw[very thick,->] (2.8,2.1) -- (3.5,2.8);
\draw[line width=6pt,yellow] (3.4,1.8) -- (3.8, 1.6);
\fill[brown!80] (3.4,1.8) circle (4pt);
\draw[decorate,decoration={coil,aspect=0,segment length=3.3pt}] (2.8,2.1) -- (3.4,1.8);
\draw (3.4,1.8) -- (5.2, 0.9);
\draw[->] (3.4,1.8) -- (4.4, 1.3);

\draw (5.2,0.9) -- (6.4, 0.1);
\draw[->] (5.2,0.9) -- (5.8,0.5);
\draw[decorate,decoration={coil,aspect=0,segment length=3.3pt}] (5.2,0.9) -- (6.9,1.5);

\draw[dashed] (1.0,0) -- (1.0,4.2);
\draw[dashed] (2,0) -- (2,3.7);
\draw[dashed] (3.8,0) -- (3.8,4.2);
\draw[dashed] (4.8,0) -- (4.8,3.7);
\draw[dashed] (6.6,0) -- (6.6,4.2);
\draw (1.5,3.5) node [align=center,anchor=base] {abs.};
\draw (2.9,3.5) node [align=center,anchor=base] {gap};
\draw (4.3,3.5) node [align=center,anchor=base] {abs.};
\draw (5.7,3.5) node [align=center,anchor=base] {gap};
\draw (2.4,3.9) node [align=center,anchor=base] {layer~1};
\draw (5.2,3.9) node [align=center,anchor=base] {layer~2};
\end{tikzpicture}
}
\caption{Sketch of a very small shower consisting of electrons (lines) and photons (wiggly lines). Colours indicate different mechanisms leading to energy deposition in layer~1. }
\label{fig:shower-sketch}
\end{figure}

A very small shower is sketched in figure~\ref{fig:shower-sketch}. The brown circle indicates the emission of a photoelectron, depositing the K-shell binding energy of the ionized atom in the first gap layer. When electrons have lost all their kinetic energy, they stop and become part of the surrounding material (indicated by a gray background). Some of the aforementioned processes produce low-energetic particles very frequently; such interactions are usually modelled  by a continuous energy loss along the entire path of the particle with a deterministic mean (yellow and gray background) and stochastic fluctuations.  Very small but frequent changes of the momentum (\emph{multiple scattering}, MSC) can be modelled by discrete changes of position and momentum, but this is not shown in figure~\ref{fig:shower-sketch}.

\begin{figure}
    \centering
        \tikzextchoice{\includegraphics{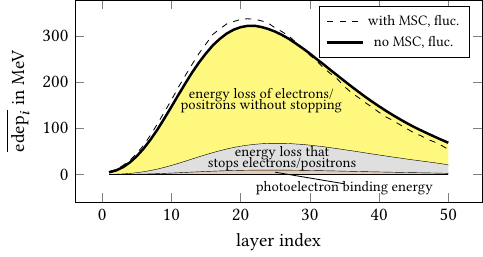}}{
    \tikzsetnextfilename{edep_primal}
    \begin{tikzpicture}
\begin{axis}[xlabel={layer index},ylabel style={align=center},ylabel near ticks,ylabel={$\samplemean{\text{edep}_{i}}$ in \si{\mega\eV}},height=5cm,width=\linewidth,legend pos=north east,ymin=-60]

\addplot[dashed] table[x index = 0, y index=1] {images/edeps-0-withmsc.dat}; 
\addlegendentry{\footnotesize with MSC, fluc.}

\addplot[name path=sum,very thick] table[x index=0, y index=9] {images/edeps.dat}; 
\addlegendentry{\footnotesize no MSC, fluc.}

\addplot[name path=border1,very thin] table[x index=0, y expr=\thisrowno{9}-\thisrowno{4}] {images/edeps.dat}; 
\addplot[name path=border2,very thin] table[x index=0, y expr=\thisrowno{9}-\thisrowno{4}-\thisrowno{2}] {images/edeps.dat}; 
\addplot[name path=border3,very thin] table[x index=0, y expr=\thisrowno{9}-\thisrowno{4}-\thisrowno{2}-\thisrowno{7}] {images/edeps.dat}; 
\addplot[name path=axis,black,samples=100,domain=1:50] {0.0};

\addplot [thick,fill=yellow, fill opacity=0.5] fill between[of=sum and border1];
\addplot [thick,fill=lightgray, fill opacity=0.5] fill between[of=border1 and border2];
\addplot [thick,fill=brown!80, fill opacity=0.5] fill between[of=border2 and border3];
\draw (axis cs:23,160) node[align=center] {\footnotesize energy loss of electrons/\\[-0.2cm] \footnotesize positrons without stopping};
\draw (axis cs:26,35) node[align=center] {\footnotesize energy loss that  \\[-0.2cm] \footnotesize stops electrons/positrons};
\draw  (axis cs:35,-30) node (photoelectron-node) {\footnotesize photoelectron binding energy};
\draw (axis cs:25,5) -- (axis cs:35,-20);
\end{axis}
\end{tikzpicture}
}
\caption{\boldmath HepEmShow-simulated average energy depositions in the 50~layers for $e=\SI[detect-weight,mode=text]{10}{\giga\eV}$, $a=\SI[detect-weight,mode=text]{2.3}{\milli\meter}$, $g=\SI[detect-weight,mode=text]{5.7}{\milli\meter}$, with multiple scattering and energy loss fluctuations enabled (dashed line) or disabled (thick line) in the simulation. A breakdown of the energy deposition without MSC and fluctuations into the main energy deposition mechanisms  is also shown (yellow, gray, brown).}
\label{fig:edep-primal}
\end{figure}

The AD outputs $f(\theta,\omega)$ analyzed in this study are given by the energy depositions $\text{edep}_i(\theta,\omega)$ in the layers $i=1,\dots,50$. Figure~\ref{fig:edep-primal} shows that disabling MSC (and energy loss fluctuations) has only a small effect on the energy depositions in our setup (as the dashed and thick lines are close to each other). The energy depositions without MSC and fluctuations are represented in figure~\ref{fig:edep-primal} as a sum of the energy deposited by the continuous energy loss of electrons and positrons (sum of yellow and gray), and the much smaller binding energies that photoelectrons leave behind (brown); other energy deposition mechanisms are mostly irrelevant in our setup. The plotted data were obtained with a particle simulation, as detailed in the next section.

\subsection{Particle Simulations}
Simulations of electromagnetic showers in material arrangements like the sampling calorimeter of section~\ref{sec:detector-geometry} can be thought of as a set of nested loops: Every iteration of the outermost \emph{event loop} is concerned with a new primary particle, and contains a \emph{stacking loop} that iterates over all particles in the resulting shower. Conceptually, each iteration of the innermost \emph{stepping loop} determines the remaining pathlength until either a volume boundary is hit or a discrete physics process happens, and then moves the particle accordingly and accounts for any effects of physics processes. 

The {\bfseries Geant4} toolkit \cite{agostinelli_geant4simulation_2003,allison_geant4_2006,allison_recent_2016} covers a wide set of particles and processes, and has a very general way to handle geometry; accordingly, it is a very complex software project with around one million lines of code, mostly written in C{\ttfamily++}. The {\bfseries G4HepEm} toolkit \cite{g4hepem-github} isolates much of Geant4's models of physics processes in electromagnetic showers; e.\,g., G4HepEm's run-time functionality includes sampling of the distance to the next discrete interaction and sampling of interaction outcomes. On the one hand, G4HepEm can be used inside of Geant4 as an alternative to Geant4's native implementation of electromagnetic physics processes.  Once the relevant material data (such as cross-sections) and other information have been pre-computed into a JSON file using separate initialization-time functionality of G4HepEm based on Geant4, G4HepEm's run-time functionality can also be used independently from Geant4, as a very compact standalone library for research and development activities in the field of HEP simulations. The {\bfseries HepEmShow} package \cite{hepemshow-github,hepemshow-doc} consists of two applications: A \emph{data generation} program using G4HepEm's initialization-time functionality and Geant4 to create the JSON file, and the main \emph{simulation} that implements event, stacking and stepping loops in the sampling calorimeter setup described above (section~\ref{sec:detector-geometry}), using physics information solely from G4HepEm's run-time functionality. HepEmShow's energy deposition results, represented by the dashed line in figure~\ref{fig:edep-primal}, are in excellent agreement with Geant4-G4HepEm's \cite{hepemshow-doc}.

\subsection{Contributions and Limitations}
In this work, we differentiate the standalone run-time part of the G4HepEm toolkit and the HepEmShow simulation application. After disabling MSC in the simulation, we successfully validate our mean pathwise derivative estimator against difference quotients, observing only a small bias. To our knowledge, this is the first time that AD has been successfully applied to a full-fledged HEP simulation. Furthermore, we demonstrate the usefulness of these derivatives in a simple gradient-based optimization study.

While this is a major step on the way towards a differentiated Geant4-scale particle simulator, our setup makes the following key simplifications:
\begin{itemize}
\item As further detailed in section~\ref{sec:results}, we have to disable MSC in the simulation. Figure~\ref{fig:edep-primal} shows that this causes a minor change in the deposited energies in the simple sampling calorimeter setup considered by us, but it can potentially become more important for other use cases of Geant4.
\item HepEmShow is made for one particular parametric geometry (figure~\ref{fig:geometry}) whereas Geant4 has a very general and flexible implementation of geometry. 
\item External electromagnetic fields, which exert forces on charged particles, are not available in HepEmShow but are fully supported in Geant4.
\item HepEmShow is only meant for simulating electromagnetic showers consisting of electrons, positrons and photons, whereas Geant4 supports all particles relevant in HEP, and many models for physics processes across wide energy ranges.
\item The present study is concerned with a limited set of AD inputs and outputs, whereas Geant4 users have very broad access to parameters and output data.
\end{itemize}

\section{Pathwise Algorithmic Differentiation}\label{sec:ad}
The HepEmShow/G4HepEm simulation code computes the averaged per-layer energy depositions $\overline{\text{edep}_i}$ ($i=1,\dots,50$) from the input data in table~\ref{tab:parameters}, notably the primary energy $e$, absorber thickness $a$ and gap thickness $g$. We have applied AD to the simulation program in order to compute averaged per-layer derivatives
\begin{equation*}
\frac{\partial\, \overline{\text{edep}_i}}{\partial\, e},
\frac{\partial\, \overline{\text{edep}_i}}{\partial\, a},
\frac{\partial\, \overline{\text{edep}_i}}{\partial\, g}.
\end{equation*}
To this end, we first applied the machine-code-based AD tool Derivgrind in a black-box fashion (Section~\ref{sec:ad-derivgrind}). After first promising observations, we switched to the operator-overloading AD tool CoDiPack (Section~\ref{sec:ad-codipack}) with a MC-specific tape size reduction technique to reduce memory usage in the reverse mode (Section~\ref{sec:ad-tapesize}). 
\subsection{Machine-Code-Based Differentiation using Derivgrind}\label{sec:ad-derivgrind}

Derivgrind \cite{aehle_forward-mode_2022} inserts AD logic into the machine code of the program to be differentiated. Therefore, only very little modification of the source code of G4HepEm and HepEmShow is required. Naturally, we had to change a few lines to indicate AD inputs ($e$, $a$, $g$) and outputs ($\text{edep}_i$) and to output the derivatives. In addition, a few G4HepEm-defined math functions like {\ttfamily G4Log} were replaced with their standard library counterparts (e.\,g.\ {\ttfamily std::log}), as their implementations perform real arithmetic via bit-wise manipulations of floating-point data in a way that might not be correctly understood by AD tools \cite{aehle2023progress}. After exploratory experiments with Derivgrind's forward mode showed encouraging results, we decided to invest the time to apply a high-performance AD tool.

\subsection{Operator-Overloading Differentiation Using CoDiPack}\label{sec:ad-codipack}

Results presented in the remainder of this study were obtained by the operator-overloading AD tool CoDiPack \cite{SaAlGauTOMS2019}. In the shape of a C{\ttfamily++} header, CoDiPack defines \emph{AD types} that behave very similar to the built-in C{\ttfamily++} floating-point types like {\ttfamily double}, but augment all real-arithmetic operations with AD logic. For maximal flexibility, we replaced most occurrences of {\ttfamily double} in the source codes of G4HepEm and HepEmShow with a type alias {\ttfamily G4double}, which we can set to {\ttfamily double}, {\ttfamily codi::RealForward} and {\ttfamily codi::RealReverse} to build non-AD, forward-mode and reverse-mode variants, respectively.  
The code is available at
\begin{center}
{\ttfamily https://github.com/SciCompKL/g4hepem/~\phantom{~}}\\
{\ttfamily https://github.com/SciCompKL/hepemshow/}
\end{center}
No type exchange has been performed
\begin{itemize}
    \item in the data generation part of HepEmShow producing the JSON file (containing pre-computed material data etc.), to avoid having to differentiate Geant4;
    \item in the JSON I/O library \cite{lohmann-json} used by the standalone part of G4HepEm -- instead, conversions between {\ttfamily double}s in the library and {\ttfamily G4double}s in G4HepEm have been added to the interface; and
    \item for variables declared as {\ttfamily constexpr}, as they must have a \emph{literal type} according to the C{\ttfamily++} standards but the CoDiPack types are not literal.
\end{itemize}

In addition to the replacements of G4HepEm-defined math functions (Section~\ref{sec:ad-derivgrind}), some manual refactoring of the source code  was necessary around uses of the {\ttfamily ?:}-operator and implicit casts to integers.  

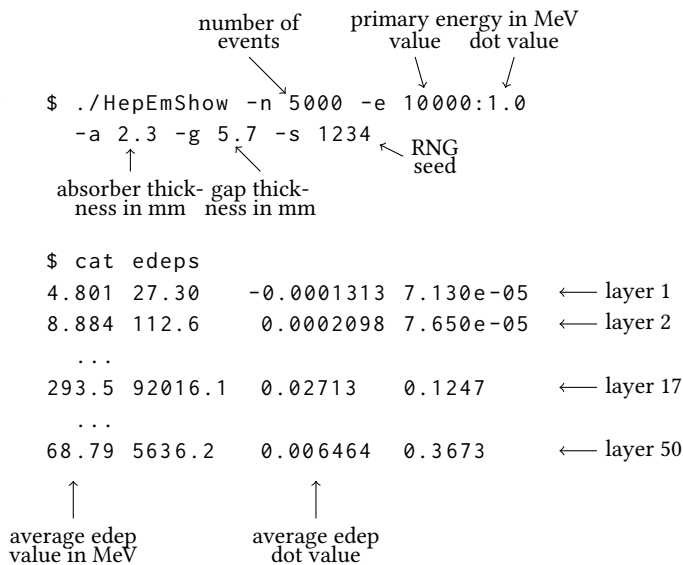
\begin{figure}
\tikzextchoice{}{\tikzexternalenable}
    \centering
    \vspace{1.2cm}
\begin{lstlisting}[basicstyle=\ttfamily,escapechar=!]
$ ./HepEmShow -n 5!\tikzmark{mark2}!000 -e 10!\tikzmark{mark3}!000:!1\tikzmark{mark1}!.0 
  -a 2!\tikzmark{mark21}!.3 -g 5!\tikzmark{mark22}!.7 -s 1234!\tikzmark{mark31}!



$ cat edeps
4.801 27.30   -0.0001313 7.130e-05 !\tikzmark{mark51}!
8.884 112.6    0.0002098 7.650e-05 !\tikzmark{mark61}!
  ...
293.5 92016.1  0.02713   0.1247    !\tikzmark{mark71}!
  ...
68!\tikzmark{mark91}!.79 563!\tikzmark{mark92}!6.2   0.00!\tikzmark{mark93}!6464  0.3!\tikzmark{mark94}!673    !\tikzmark{mark81}!
!\quad!
!\quad!
\end{lstlisting}
\begin{tikzpicture}[remember picture,overlay]
\draw[<-,shorten <=0.3cm] (pic cs:mark2) -- +(-0.7,0.7) node[above,align=center]{number of \\[-0.14cm] events};
\draw[<-,shorten <=0.3cm] (pic cs:mark3) -- +(-0.2,0.7) node[above] (e-value) {value};
\draw[<-,shorten <=0.3cm] (pic cs:mark1) -- +(0.2,0.7) node[above] (e-dotvalue) {dot value};
\draw[<-,shorten <=0.1cm] (pic cs:mark21) -- +(0.,-0.4) node[below,align=center] {absorber thick-\\[-0.14cm] ness in \si{\milli\meter}};
\draw[<-,shorten <=0.1cm] (pic cs:mark22) -- +(0.4,-0.4) node[below,align=center] {gap thick-\\[-0.14cm] ness in \si{\milli\meter}};
\draw[<-,shorten <=0.1cm] (pic cs:mark31) -- +(0.4,-0.2) node[right,align=center] {RNG \\[-0.14cm] seed};
\draw ($0.5*(e-value)+0.5*(e-dotvalue)$) node[above] {primary energy in \si{\mega\eV}};

\draw [<-] ($(pic cs:mark51)+(0.2,0.1)$) -- +(0.5,0) node[right]{layer~1};
\draw [<-] ($(pic cs:mark61)+(0.2,0.1)$) -- +(0.5,0) node[right]{layer~2};
\draw [<-] ($(pic cs:mark71)+(0.2,0.1)$) -- +(0.5,0) node[right]{layer~17};
\draw [<-] ($(pic cs:mark81)+(0.2,0.1)$) -- +(0.5,0) node[right]{layer~50};
\draw [<-] ($(pic cs:mark91)+(0.0,-0.3)$) -- +(0,-0.5) node[below,align=center]{average edep\\[-0.14cm] value in \si{\mega\eV}};
\draw [<-] ($(pic cs:mark93)+(0.0,-0.3)$) -- +(0,-0.5) node[below,align=center]{average edep\\[-0.14cm] dot value};
\end{tikzpicture}
\caption{User interface of the differentiated HepEmShow application in the forward mode. Dot values of inputs are specified in the command line interface (here, shown for {\ttfamily -e}). Dot values of outputs are written to a file {\ttfamily edeps}. }
\label{fig:diff-hepemshow-ui}
\tikzextchoice{}{\tikzexternalenable}
\end{figure}

We have extended HepEmShow's I/O to allow the user to specify the AD inputs and outputs. As shown in figure~\ref{fig:diff-hepemshow-ui}, in the forward mode, the user can supply dot values of the primary energy $e$, absorber thickness $a$, and gap thickness $g$ and access dot values of the average edeps.  Reverse-mode HepEmShow requires an additional command-line argument {\ttfamily -b} with the adjoint values of the mean edeps in all layers, separated by colons, and output the adjoint values of $e$, $a$ and $g$ in a file.

For other variables used by HepEmShow, adding them as AD inputs and outputs would likely be straightforward. However, as we have not differentiated the initialization-time functionality (which uses Geant4), it is not possible at the moment to declare Geant4-internal data (e.\,g.\ cross-section tables) as AD inputs.

\subsection{Reduction of the Tape Size}\label{sec:ad-tapesize}
In the reverse mode of AD, operator-overloading AD tools record a \emph{tape} data structure storing the real-arithmetic evaluation graph from the inputs to the outputs. For long-running programs, the tape size might exceed the amount of available memory. In our case, the recording of a single event loop iteration occupies roughly around \SI{250}{\mega\byte} of memory on the tape (measured for $e=\SI{10}{\giga\eV}$, $a=\SI{2.3}{\milli\meter}$, $g = \SI{5.7}{\milli\meter}$). As all event loop iterations run independently from each other, only a single iteration must be stored at a time, and the corresponding section of the tape can be evaluated and cleared at the end of each iteration to limit the tape size \cite{Hascoet2002}.

As source transformation tools have access to the entire source code of the function to be differentiated, they can generally use smaller tapes and apply more advanced code optimizations.  As it is possible to compile the HepEmShow  simulation and its G4HepEm dependency in a single translation unit, it would be worthwhile to investigate how compiler-based source transformation AD tools such as Clad~\cite{Vassilev_Clad} perform on the code base.

\section{Results}\label{sec:results}

\subsection{Stochastic Noise with and without Multiple Scattering}\label{sec:noise}
We first take a look at how the energy deposition depends on the primary particle energy $e$ without AD, in order to be able to explain our findings with AD in the next section~\ref{sec:variance-bias}.

\begin{figure}
    \centering
    \tikzextchoice{\includegraphics{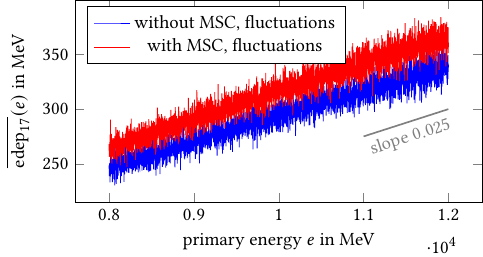}}{
    \tikzsetnextfilename{stochastic_noise_large}
    \begin{tikzpicture}
\begin{axis}[xlabel={primary energy $e$ in \si{\mega\eV}},ylabel style={align=center},ylabel near ticks,ylabel={$\samplemean{\text{edep}_{17}}(e)$ in \si{\mega\eV}},height=5cm,width=\linewidth,legend pos=north west]
\addplot[blue] table[x index=0, y index=17,] {images/slider/outputs-nomsc.dat}; 
\addplot[red] table[x index=0, y index=17,] {images/slider/outputs-msc.dat}; 
\draw [gray,thick] (axis cs: 11000,275) -- (axis cs: 12000,300) node[midway,sloped,below]{slope $0.025$};
\addlegendentry{without MSC, fluctuations}
\addlegendentry{with MSC, fluctuations}
\end{axis}
\end{tikzpicture}
}
\caption{\boldmath Dependency of the simulated mean energy deposition in layer~17 on the primary energy $e$. For every point in this plot, $N=100$ events were simulated using the same random seed. Figure~\ref{fig:stochastic-noise-small} zooms into this plot to see if these ``noisy'' functions are differentiable and if their derivatives match the large-scale slope of $0.025$.}
\label{fig:stochastic-noise-large}
\end{figure}
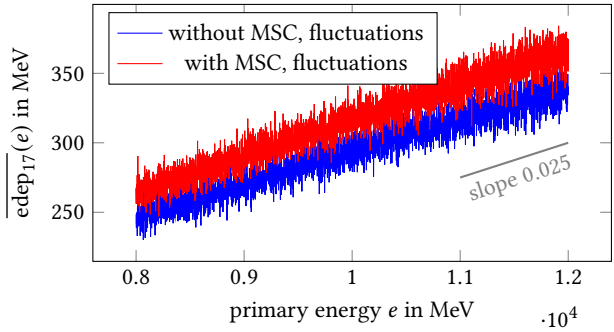

{\bfseries Large scale.} Figure~\ref{fig:stochastic-noise-large} shows the simulated mean energy deposition in layer~17, $\samplemean{\text{edep}_{17}}$, averaged over $N=100$ events per run, as a function of primary particle energy $e$. Each of the 4001 data points between $e=\SI{8000}{\mega\eV}$ and \SI{12000}{\mega\eV} was produced by a separate run of HepEmShow, always using the same initial random seed. The experiment has been conducted with the full set of electromagnetic processes available in G4HepEm (red), and with a simplified setup that had MSC and energy loss fluctuations disabled (blue).

The number of $N=100$ simulated events for figure~\ref{fig:stochastic-noise-large} is very small, so the standard deviation of the mean \eqref{eq:average} is rather large, causing the clearly visible stochastic noise. This is expected: If $e$ is perturbed even very slightly, the control flow in the simulator is likely to change at some point, making a different number of calls to the RNG and thus leaving it in a different state for the subsequent execution, which is therefore entirely uncorrelated even though the same RNG seed has been used initially \cite{Aehle_2023}. Choosing a higher $N$ reduces the amplitude of the stochastic noise, but does not eliminate it.

Despite the noise, figure~\ref{fig:stochastic-noise-large} shows a clear large-scale trend, with $\samplemean{\text{edep}_{17}}(e)$ rising, in both setups, approximately linearly by \SI{100}{\mega\eV} over the entire range of $e$ spanning \SI{4000}{\mega\eV}. Thus, the derivative $(\EE\,\text{edep}_{17})'$ of the expected energy deposition at $e=\SI{10000}{\mega\eV}$ can be estimated as 
\begin{equation}\label{eq:sought-derivative}
(\EE\,\text{edep}_{17})' \dotequal \tfrac{\partial }{\partial e}\left[\EE_\omega{\text{edep}_{17}}(e,\omega)\right] \approx \frac{\SI{100}{\mega\eV}}{\SI{4000}{\mega\eV}} = 0.025.
\end{equation}
This \emph{large-scale} slope is what is relevant for e.\,g.\ optimization purposes, so this is what we want to compute. For validation purposes, we approximate the large-scale derivative using difference quotients similar to the right-hand side of \eqref{eq:sought-derivative}, taking care that a sufficiently large number of events is simulated as difference quotients are poorly conditioned.

When we apply AD to a code computing $\samplemean{\text{edep}_{17}}$ , we obtain the floating-point accurate, \emph{local} slope $\samplemean{\text{edep}_{17}'}$ of the algorithm implemented in the code. To read this local slope from the plot, we have to zoom in.

\begin{figure}
    \begin{flushright}
        \tikzextchoice{\includegraphics{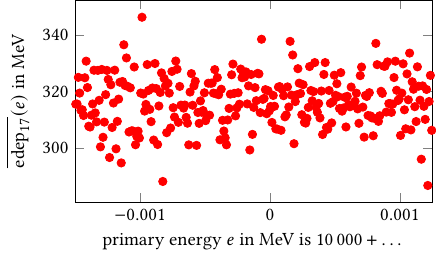}}{
    \tikzsetnextfilename{stochastic_noise_small_1}
    \begin{tikzpicture}
\begin{axis}[xlabel={primary energy $e$ in \si{\mega\eV} is $10\,000+\dots$},ylabel style={align=center},ylabel near ticks,ylabel={$\samplemean{\text{edep}_{17}}(e)$ in \si{\mega\eV}},height=5cm,width=0.9\linewidth,xmin=-0.0015,xmax=0.00125,xtick={-0.001,0,0.001},xticklabels={$-0.001$,$0$,$0.001$},scaled x ticks=false]
\addplot[red,mark=*,only marks] table[x index=0, y index=17,] {images/slider/outputs-zoom-msc.dat}; 
\end{axis}
\end{tikzpicture}
}
\\[0.3cm]
    \tikzextchoice{\includegraphics{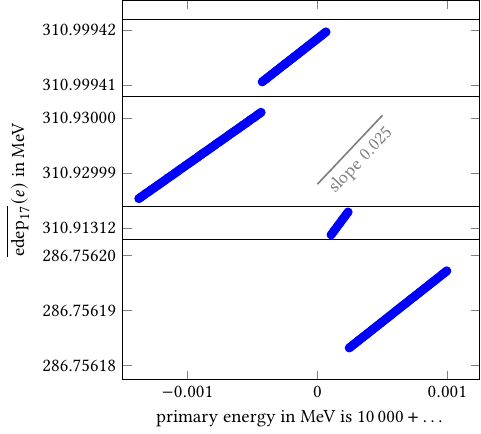}}{
    \tikzsetnextfilename{stochastic_noise_small_2}
\begin{tikzpicture}
\input{images/slider/processed-outputs-nomsc.tex} 
\end{tikzpicture}
}
\end{flushright}
\caption{\boldmath Zoom into figure~\ref{fig:stochastic-noise-large}, showing a much smaller range of $e$. Again, each point represents a HepEmShow simulation of $N=100$ events, always using the same random seed. The energy deposition computed with the full set of physics processes still appears noisy (top). With multiple scattering and energy loss fluctuations disabled, however, the averaged energy deposition is a piecewise differentiable function of the primary energy, and its derivative (i.\,e.\ the slope of the segments) approximately matches the large-scale slope.}
\label{fig:stochastic-noise-small}
\end{figure}

{\bfseries Small scale.} Figure~\ref{fig:stochastic-noise-small} shows $\samplemean{\text{edep}_{17}}(e)$ plotted over a much more narrow interval, again using the same seed for all runs of HepEmShow. For the full physics setup, we observe the same noisy behavior (top figure), even if we zoomed in further. With MSC and energy loss fluctuations turned off, however, the function is clearly piecewise differentiable (bottom figure). This qualitative difference is very important for AD, as it allows us to confirm that the slopes of the differentiable segments (which is what pathwise AD computes) are close to the large-scale slope of about $0.025$ as determined in \eqref{eq:sought-derivative} which we want to compute. There is still more than one jump per \si{\kilo\eV} on the horizontal axis, due to discrete randomness and decorrelating RNG states as mentioned above. These jumps are much larger in magnitude than the differentiable evolution in between, and they are responsible for the noise visible in figure~\ref{fig:stochastic-noise-large}. However, the differentiable evolution in between the jumps already accounts for (approximately) the entire large-scale evolution.

In fact, it is not necessary to disable energy loss fluctuations; a plot in the style of figure~\ref{fig:stochastic-noise-small}, with only multiple scattering turned off, has more discontinuities but still clearly visible increasing differentiable segments.

{\bfseries Summary.} The qualitative analysis conducted in this section for a single layer indicates that after disabling MSC, there is only a small difference between
\begin{itemize}
    \item the large-scale derivative $(\EE\,\text{edep}_{17})'$ required for applications and approximated by difference quotients, and
    \item the local derivative $\samplemean{\text{edep}_{17}'}$, or $\EE(\text{edep}_{17}')$ in the limit of many simulated events ($N_\text{diff}\to\infty$), computed by AD without special care for randomness (thus treating random numbers as constants). 
\end{itemize}
In the next section~\ref{sec:variance-bias}, we study this hypothesis quantitatively and in more generality, looking at algorithmic derivatives of energy depositions in all layers with respect to $e$, $a$ and $g$.

\subsection{Variance and Bias of Pathwise Algorithmic Derivatives}\label{sec:variance-bias}

In this section, we collect results obtained with our CoDiPack-differentiated version of HepEmShow/G4HepEm (sections~\ref{sec:ad-codipack}, \ref{sec:ad-tapesize}). 

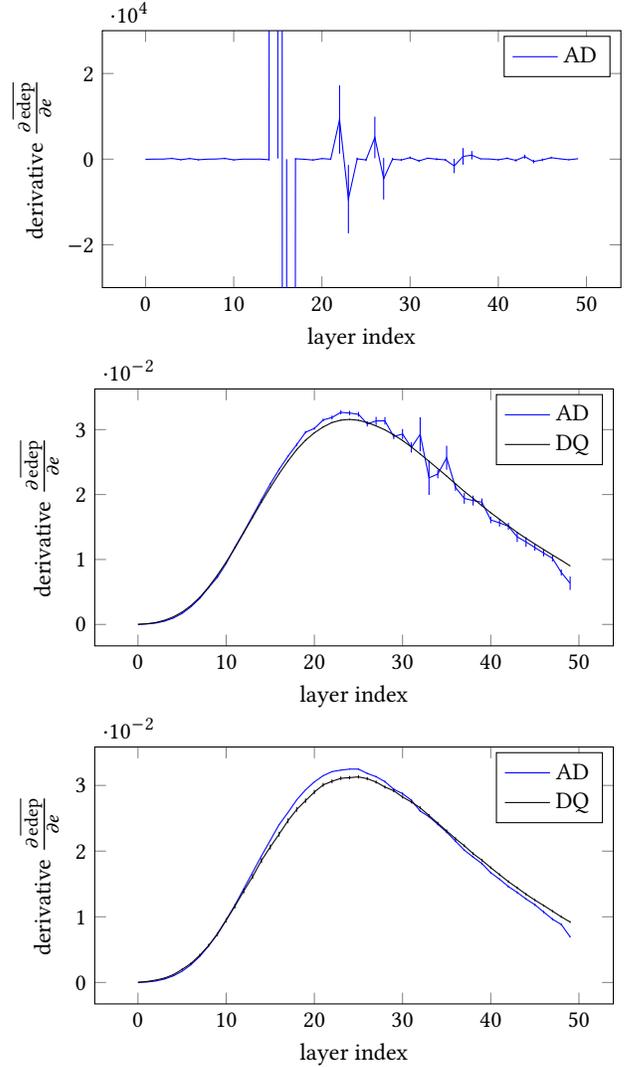
\begin{figure}
    \centering
        \tikzextchoice{\includegraphics{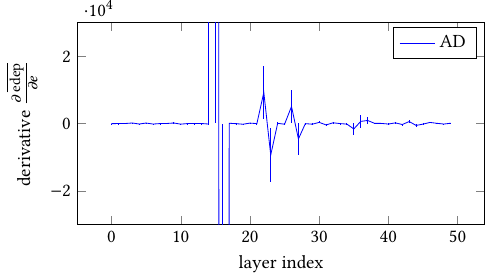}}{
    \tikzsetnextfilename{d_edep_d_primaryenergy_1}
    \begin{tikzpicture}
\begin{axis}[ylabel={derivative $\tfrac{\partial\,\samplemean{\text{edep}}}{\partial e}$},ylabel style={align=center},ylabel near ticks,xlabel={layer index},height=5cm,error bars/y dir=both,error bars/y explicit,error bars/error bar style={thin},error bars/error mark=none,width=\linewidth,ymin=-30000,ymax=30000]
\addplot[blue] table[x index=0, y index=1,y error index=2] {images/no-simplifications.dat};
\addlegendentry{AD}
\end{axis}
\end{tikzpicture}
}
\tikzextchoice{\includegraphics{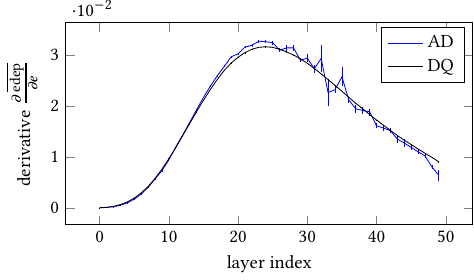}}{
    \tikzsetnextfilename{d_edep_d_primaryenergy_2}
\begin{tikzpicture}
\begin{axis}[ylabel={derivative $\tfrac{\partial\,\samplemean{\text{edep}}}{\partial e}$},ylabel style={align=center},ylabel near ticks,xlabel={layer index},height=5cm,error bars/y dir=both,error bars/y explicit,error bars/error bar style={thin},error bars/error mark=none,width=\linewidth]
\addplot[blue] table[x index=0, y index=1,y error index=2] {images/with-fluctuation.dat};
\addlegendentry{AD}
\addplot[black] table[x index=0, y index=3,y error index=4] {images/with-fluctuation.dat};
\addlegendentry{DQ}
\end{axis}
\end{tikzpicture}
}
\tikzextchoice{\includegraphics{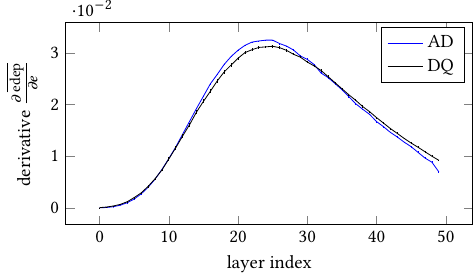}}{
    \tikzsetnextfilename{d_edep_d_primaryenergy_3}
\begin{tikzpicture}
\begin{axis}[ylabel={derivative $\tfrac{\partial\,\samplemean{\text{edep}}}{\partial e}$},ylabel style={align=center},ylabel near ticks,xlabel={layer index},height=5cm,error bars/y dir=both,error bars/y explicit,error bars/error bar style={thin},error bars/error mark=none,width=\linewidth]
\addplot[blue] table[x index=0, y index=1,y error index=2] {images/newlong.dat};
\addlegendentry{AD}
\addplot[black] table[x index=0, y index=3,y error index=4] {images/newlong.dat};
\addlegendentry{DQ}
\end{axis}
\end{tikzpicture}
}
    \caption{\boldmath Algorithmic derivative of the mean edep in the calorimeter layers with respect to the primary energy $e$ (blue), and the corresponding difference quotients (black). Error bars indicate 68\,\%-confidence intervals (i.\,e.\ plus/minus one standard deviation). Top: Default configuration of G4HepEm with all physics processes. Middle: All physics processes except for multiple scattering. Bottom: More samples and smaller interval for the difference quotient.}
    \label{fig:d-edep-d-primaryenergy}
\end{figure}

{\bfseries Pathwise derivatives of the full simulation code including MSC are noisy.}
Figure~\ref{fig:d-edep-d-primaryenergy} shows the mean pathwise forward-mode algorithmic derivative of the simulated energy deposition $\samplemean{\text{edep}_i(e)}$ in all the calorimeter layers $i=0, \dots, 49$, with respect to the initial kinetic energy $e$ of the primary particles, at $e=\SI{10}{\giga\eV}$. For the top plot, 24\,M events were simulated using the full list of physics processes. Mean pathwise derivatives $\samplemean{\text{edep}'_i(e)}$ of the code seem to have a very large variance and deviate by orders of magnitudes from the value of $0.025$ suggested by \eqref{eq:sought-derivative}. 
Averaging over many more events might reduce noise, but as the number of events would need to rise by a factor of $10^{12}$ to bring a standard deviation of the order of $10^{4}$ down to the order of $10^{-2}$, this is not feasible in practice.

It should be noted that this observation does not imply that the physical phenomenon of MSC itself would be inherently non-differentiable. We can only infer that G4HepEm's algorithm implementing the Urban MSC model \cite{Urban:2006kd} has noisy algorithmic derivatives. This could be related to the often-heard statement that ``black-box'' differentiation of iterative numerical algorithms may compute wrong derivatives \cite{doi:10.1080/10556789208805503} and knowledge on the mathematical structure behind them should thus be included into the AD implementation. For the remainder of this study, however, we disable MSC in the simulation, and leave the development of an AD-friendly MSC model to further research. 

{\bfseries Disabling MSC leads to low variance and bias for pathwise derivatives.} 
The middle plot of figure~\ref{fig:d-edep-d-primaryenergy} shows the averaged result of 24\,M AD runs at $e=\SI{10}{\giga\eV}$ with MSC disabled in the simulation. Additionally, 24\,M primal runs without MSC at $e=\SI{9.9}{\giga\eV}$ and $e=\SI{10.1}{\giga\eV}$ were conducted to compute a central difference quotient (DQ) that approximates the large-scale slope ($\EE\,\text{edep}_i(e))'$. Both plots match very well. Thus, disabling multiple scattering is the key algorithmic change that allows us to obtain algorithmic derivatives with a sufficiently low variance and an expected value close to the numerical derivatives. 

{\bfseries The bias with respect to difference quotients is around~5\,\%.} The bottom plot of figure~\ref{fig:d-edep-d-primaryenergy} has been created with 864\,M samples to decrease the stochastic error, and a more narrow interval {9.995\dots10.005}\,\si{\giga\eV} for the difference quotient to decrease the numerical truncation error, again with MSC disabled. We observe a statistically highly significant but low deviation of the mean pathwise derivative approximating $\EE (\text{edep}_i')$ from the difference quotients approximating $(\EE\,\text{edep}_i)'$. Except for the first and last few layers, the relative error of the derivatives is around \SI{5}{\percent}.

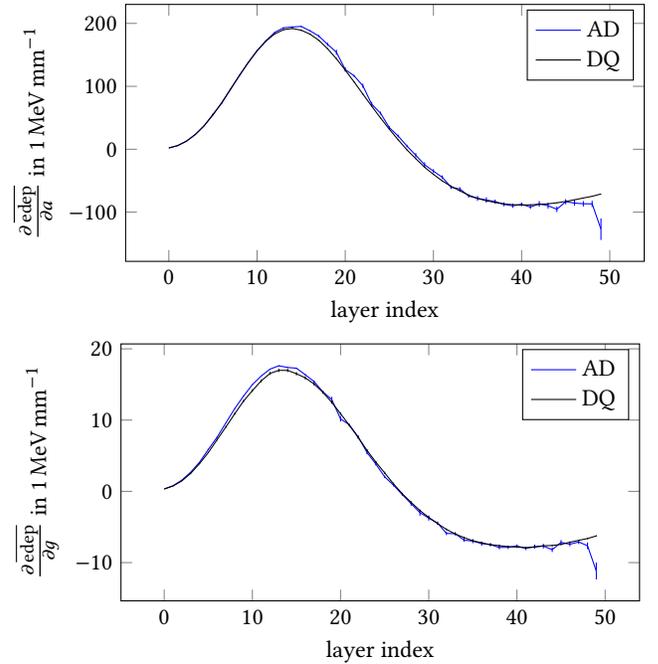
\begin{figure}
    \centering
    \tikzextchoice{\includegraphics{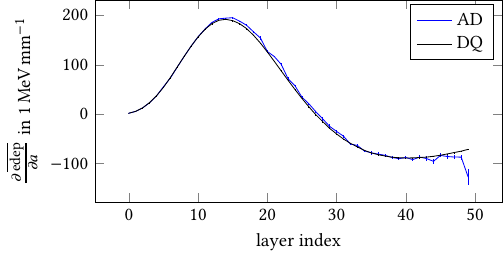}}{
    \tikzsetnextfilename{d_edep_d_thickness_1}
    \begin{tikzpicture}
\begin{axis}[ylabel={$\tfrac{\partial\,\samplemean{\text{edep}}}{\partial a}$ in \SI{1}{\mega\eV\per\milli\meter}},ylabel near ticks,ylabel style={align=center},xlabel={layer index},height=5cm,error bars/y dir=both,error bars/y explicit,error bars/error bar style={thin},error bars/error mark=none,width=\linewidth]
\addplot[blue] table[x index=0, y index=1,y error index=2] {images/wrt-thickness.dat};
\addlegendentry{AD}
\addplot[black] table[x index=0, y index=3,y error index=4] {images/wrt-thickness.dat};
\addlegendentry{DQ}
\end{axis}
\end{tikzpicture}
}
\tikzextchoice{\includegraphics{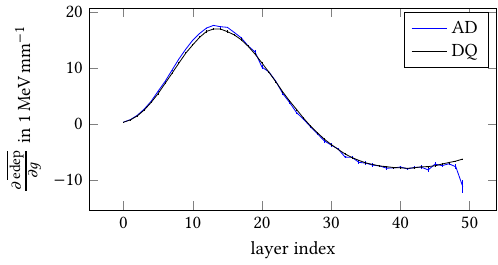}}{
    \tikzsetnextfilename{d_edep_d_thickness_2}
    \begin{tikzpicture}
\begin{axis}[ylabel={$\tfrac{\partial\,\samplemean{\text{edep}}}{\partial g}$ in \SI{1}{\mega\eV\per\milli\meter}},ylabel near ticks,ylabel style={align=center},xlabel={layer index},height=5cm,error bars/y dir=both,error bars/y explicit,error bars/error bar style={thin},error bars/error mark=none,width=\linewidth]
\addplot[blue] table[x index=0, y index=1,y error index=2] {images/wrt-gap-thickness.dat};
\addlegendentry{AD}
\addplot[black] table[x index=0, y index=3,y error index=4] {images/wrt-gap-thickness.dat};
\addlegendentry{DQ}
\end{axis}
\end{tikzpicture}
}
    \caption{\boldmath Algorithmic derivative of the edep with respect to the absorber thickness $a$ (top) and gap thickness $g$ (bottom).}
    \label{fig:d-edep-d-thickness}
\end{figure}

{\bfseries Similar observations can be made for derivatives w.\,r.\,t.\ layer thicknesses.} Figure~\ref{fig:d-edep-d-thickness} shows that algorithmic derivatives of the energy deposition with respect to the absorber and gap thicknesses as well have a sufficiently low variance and bias (w.\,r.\,t.\ difference quotients) when MSC is turned off.

{\bfseries Algorithmic and numeric derivatives deviate much more for individual edep mechanisms.}
While the mean pathwise derivatives of the total energy depositions $\text{edep}_i$ are close to the large-scale derivatives approximated by difference quotients, as described above, this does not hold on the level of individual mechanisms to register energy deposition in the simulation code. 

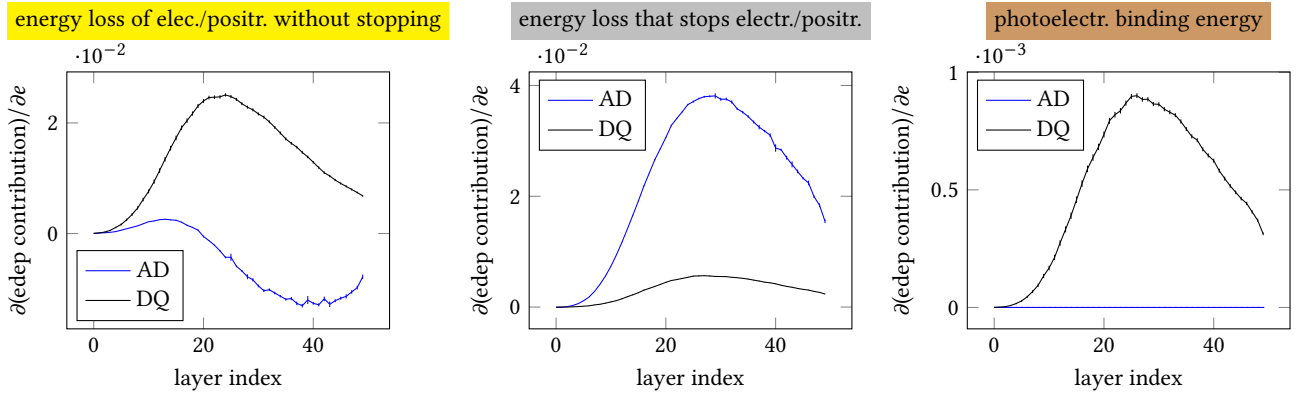
\begin{figure*}
    \centering
    \tikzextchoice{\includegraphics{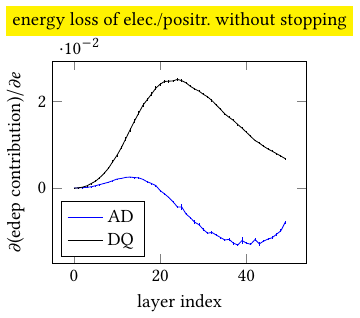}}{
    \tikzsetnextfilename{d_edep_d_primaryenergy_breakdown_1}
    \begin{tikzpicture}
\begin{axis}[ylabel={$\partial \text{(edep contribution)} / \partial e$},ylabel style={align=center},ylabel near ticks,xlabel={layer index},height=5cm,error bars/y dir=both,error bars/y explicit,error bars/error bar style={thin},error bars/error mark=none,width=0.33\linewidth,title style={align=center},title={\colorbox{yellow}{energy loss of elec./positr.\ without stopping}\\ \vspace{-0.1cm}},legend pos=south west]

\addplot[blue] table[x index=0, y index=15,y error index=16] {images/edepd_breakdown.dat};
\addlegendentry{AD}
\addplot[black] table[x index=0, y index=13,y error index=14] {images/edepd_breakdown.dat};
\addlegendentry{DQ}
\end{axis}
\end{tikzpicture}
}
\tikzextchoice{\includegraphics{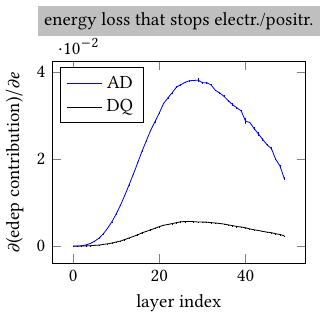}}{
    \tikzsetnextfilename{d_edep_d_primaryenergy_breakdown_2}
    \begin{tikzpicture}
\begin{axis}[ylabel={$\partial \text{(edep contribution)} / \partial e$},ylabel style={align=center},ylabel near ticks,xlabel={layer index},height=5cm,error bars/y dir=both,error bars/y explicit,error bars/error bar style={thin},error bars/error mark=none,width=0.33\linewidth,title style={align=center},title={\colorbox{lightgray}{energy loss that stops electr./positr.}\\ \vspace{-0.1cm}},legend pos=north west]

\addplot[blue] table[x index=0, y index=7,y error index=8] {images/edepd_breakdown.dat};
\addlegendentry{AD}
\addplot[black] table[x index=0, y index=5,y error index=6] {images/edepd_breakdown.dat};
\addlegendentry{DQ}
\end{axis}
\end{tikzpicture}
}
    \tikzextchoice{\includegraphics{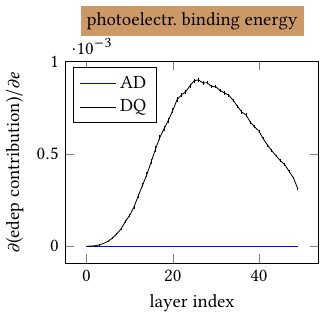}}{
    \tikzsetnextfilename{d_edep_d_primaryenergy_breakdown_3}
    \begin{tikzpicture}
\begin{axis}[ylabel={$\partial \text{(edep contribution)} / \partial e$},ylabel style={align=center},ylabel near ticks,xlabel={layer index},height=5cm,error bars/y dir=both,error bars/y explicit,error bars/error bar style={thin},error bars/error mark=none,width=0.33\linewidth,title style={align=center},title={\colorbox{brown!80}{photoelectr.\ binding energy}\\ \vspace{-0.1cm}},legend pos=north west]

\addplot[blue] table[x index=0, y index=27,y error index=28] {images/edepd_breakdown.dat};
\addlegendentry{AD}
\addplot[black] table[x index=0, y index=25,y error index=26] {images/edepd_breakdown.dat};
\addlegendentry{DQ}
\end{axis}
\end{tikzpicture}
}
    \caption{\boldmath Breakdown of $\partial\,\text{edep}_i / \partial\,e$ into the dominating energy deposition mechanisms. }
    \label{fig:d-edep-d-primaryenergy-breakdown}
\end{figure*}

We have used figure~\ref{fig:edep-primal} to illustrate that most of the energy deposition comes from continuous energy loss, followed by the binding energy of photoelectrons. In fact, G4HepEm registers continuous energy loss at two main places in the code: As a side action next to another physical process or a change of volumes in the geometry (indicated in yellow) and as the sole action if it uses up all the remaining kinetic energy of the particle (indicated in gray).   Figure~\ref{fig:d-edep-d-primaryenergy-breakdown} shows the derivatives of these three terms, with respect to the primary energy $e$ again. Interestingly, algorithmic and numeric derivatives do not match.

To understand the algorithmic derivatives of the two energy loss contributions, let us imagine that the incoming electron in figure~\ref{fig:shower-sketch} had an infinitesimally higher initial kinetic energy. This would allow the primary and the secondary electrons to travel infinitesimally further before they stop, making the gray segments longer and their energy deposition higher. Thus, this mechanism contributes most to the algorithmic derivative of the energy deposition.

Regarding the difference quotients, we have to imagine a small but non-infinitesimal increase in the initial kinetic energy. As before, gray segments become longer, but one of the secondary electrons may now have enough energy to reach the next layer, and the energy loss would become a side action. Therefore, difference quotients mainly see an increase in energy loss that does not stop the particle.

We should note that the distinction between the two mechanisms to register continuous energy loss comes from modelling and coding considerations and is not rooted in physics. Our observation that algorithmic and numerical derivatives deviate heavily for the two individual mechanisms, even though they approximately match for the sum, shows that care should be taken to only declare physically meaningful data as AD outputs.

Concerning the deposition of binding energies in photoelectric effect events, difference quotients register an increase that could be caused by more events taking place, and/or an increasing probability of elements with higher binding energies to be selected as the ionized atom from the material. Both types of dependencies have the structure of $f_2$ in figure~\ref{fig:pathwise-derivative-model}, and are thus not seen by pathwise algorithmic derivatives, which are therefore zero. This illustrates that we cannot expect pathwise derivatives to perfectly match the numerical derivatives. We should note that the deviation stated here is not equal to the observed bias of around \SI{5}{\percent}.

{\bfseries Performance Measurements.} Table~\ref{tab:runtimes} shows the runtime and memory consumption of a HepEmShow simulation of 10\,000 electrons, in terms of user time and maximum resident set size measured on an exclusive \SI{2.6}{\giga\hertz} Intel Xeon Gold 6126 node at the Elwetritsch cluster of the University of Kaiserslautern-Landau. 

Forward-mode and reverse-mode AD using CoDiPack slow down the program by factors of around 1.8 and 5.4, respectively. Note that a single reverse-mode AD run computes an entire gradient, so reverse-mode AD is faster than central difference quotients if a gradient with respect to more than two input variables is sought. 

Memory consumption increased slightly in the forward mode because CoDiPack's forward-mode type has twice the size of a {\ttfamily double}. In the reverse mode, the tape occupies a significant but perfectly manageable amount of memory, which grows with the primary energy.

\begin{table}
    \centering
    \begin{tabular}{lcccccc}
         \toprule
         Primary  & \multicolumn{2}{c}{\SI{5}{\giga\eV}}  & \multicolumn{2}{c}{\SI{10}{\giga\eV}} & \multicolumn{2}{c}{\SI{20}{\giga\eV}}  \\
         energy & time & mem. & time & mem. & time & mem. \\
         
         \midrule
         primal  & 84 & 5.7 & 163 & 5.6 & 320 & 5.7 \\
         \midrule
         forward  & 147 & 5.9 & 287 & 5.9 & 558 & 5.9 \\
         mode & ($\times 1.8$) & & ($\times 1.8$) & & ($\times 1.7$)\\
         \midrule
         reverse  & 452 & 111 & 867 & 195 & 1662 & 284 \\
         mode & ($\times 5.4$) & & ($\times 5.3$) & & ($\times 5.2$)\\
         \bottomrule
    \end{tabular}
    \caption{\boldmath Runtime (in seconds) and memory (in MB) required to simulate 10\,000 electron events for $e=5, 10, \SI[detect-weight,mode=text]{20}{\giga\eV}$.}
    \label{tab:runtimes}
\end{table}

\subsection{Optimization Using Averages of Pathwise Derivatives}\label{sec:optimization}

This section deals with an applications of pathwise algorithmic derivatives for gradient-based optimization.

The \emph{gradient descent algorithm} attemps to find the minimizer $\theta^* \in \RR^n$ of a loss function $L: \RR^n \to \RR$, starting from an initial guess $\theta^{(0)} \in \RR^n$, by iteratively computing better and better ``candidate minimizers'' $\theta^{(1)}$, $\theta^{(2)}$, \dots via
\begin{equation}\label{eq:gradient-descent}
    \theta^{(k+1)}_j = \theta^{(k)}_j - d_j^{(k)} \cdot \tfrac{\partial L}{\partial \theta_j}(\theta^{(k)}).
\end{equation}
The factors $d_j^{(k)}$ are called  \emph{step-sizes} or \emph{learning rates}, and may be fixed or computed adaptively. When stochastic estimates are used instead of the actual gradient, the scheme is known as \emph{stochastic gradient descent} (SGD). In machine learning, stochastic estimates of loss function gradients typically result from computing the loss on a randomly selected subset of the training data instead of the entire data. In our case, outputs of the MC simulation, and hence their derivatives, are stochastic already by definition.  Deviations of the estimated derivatives from the true values steer the optimizer into a less ideal direction, but it can still arrive at the minimum, maybe with a larger number of steps.

\textbf{Automated Design of Scientific Instruments.} To demonstrate that the stochastic and biased pathwise AD gradient estimator can indeed be useful for optimization, we have designed the following simple parameter identification problem. The parameters in table~\ref{tab:parameters} have been used to simulate a target edep distribution $\samplemean{\text{edep}_i}$ across the layers $i=0, \dots, 49$ of the calorimeter, shown in figure~\ref{fig:edep-primal}. From this target edep distribution, we wish to infer $e^*=\SI{10}{\giga\eV}$ and $a^* = \SI{2.3}{\milli\meter}$, assuming that we only know the other parameters in table~\ref{tab:parameters}. The task of identifying a primary energy value $e^*$ that leads to a prescribed energy deposition curve is a model problem for %
applications where the position of physical interactions should be controlled, e.\,g.\ for experiment design or in radiation therapy planning.
Derivatives with respect to a geometric parameter $a^*$ could be useful for detector optimization problems.  

To identify $e^*$ and $a^*$, we have to search for the minimizer of the loss function $L$ given by the squared error of the resulting edep distribution,
\begin{equation}\label{eq:loss-L}
    L(e,a) = \sum_{i=0}^{49} \left(\samplemean{\text{edep}_i}(e,a) - \samplemean{\text{edep}_i}(e^*,a^*)\right)^2.
\end{equation}

Figure~\ref{fig:optimization} shows 16~paths of the stochastic gradient descent scheme across the loss landscape of $L$. We have chosen a step-size of $1$ for $e$ and $10^{-7}\,\si{\milli\meter\squared\per\mega\eV\squared}$ for $a$ to account for their different units and orders of magnitude, and estimate the gradient using 1\,k events in each step, for 350 steps. Starting from $e^{(0)}=\SI{22}{\giga\eV}$ and $a^{(0)}=\SI{3}{\milli\meter}$, the SGD optimizer robustly converges to the minimizer $(e^*,a^*)$. There is room for further investigation of optimal choices of such hyperparameters; e.\,g., the optimization succeeds even with only 100 events per step.

\section{Conclusion and Outlook}\label{sec:conclusion}
In this work, we have successfully applied AD to a Monte-Carlo simulation of electromagnetic showers in a sampling calorimeter, in order to compute pathwise derivatives of the energy depositions with respect to the energy of the primary particles and the thicknesses of the layers. The simulation models all the relevant physics processes, while the detector geometry has been kept rather simple. Applying AD to the code without any algorithmic changes led to algorithmic derivatives of very high variance, but the only problem seems to be that black-box pathwise AD is not the right tool to differentiate the algorithm used to model multiple scattering in G4HepEm. With multiple scattering disabled, variances of algorithmic derivatives are sufficiently low and their means are close to the (numerical) derivatives of the average energy depositions, with a deviation of about \SI{5}{\percent}. Errors of this magnitude may be perfectly acceptable when the derivatives are used for gradient-based optimization, as demonstrated by a simple parameter identification study.

In order to scale our encouraging result to the full generality of Geant4, we propose the following next steps:

\begin{itemize}
\item It could be worthwhile to apply a high-performance AD tool to the Geant4 codebase, in order to try to reproduce the findings of our present work with Geant4's G4HepEm physics process. This would allow to see if Geant4's very general implementation of geometry is an obstacle for AD, and if not, allow to consider many different detector layouts.
\item Subsequently, Geant4's G4HepEm physics process could be replaced by the native Geant4 electromagnetic physics processes.
\item Additional efforts should be dedicated to analyze and mitigate the incompatibility of AD with multiple scattering, potentially by creating an AD-friendly MSC model.
\item One could then enable uniform and non-uniform electromagnetic fields in the simulation to check their compatibility with AD.
\item To conclude physics generalizations, it would be interesting to include other particles and, in particular, enable hadronic processes. 
\item At some point, it may become necessary to go beyond mean pathwise derivatives, and employ and improve differentiable and probabilistic programming tooling to account for discrete randomness.
\item In particular, systematic efforts should be dedicated to source transformation AD tools to enable the above workflows. 
\item Once any pre- and postprocessing software is differentiated as well, algorithmic derivatives can be used to efficiently optimize actual experiment designs in their planning phase.
\end{itemize}

\section{Author Contributions}
{\bfseries Max Aehle}: Conceptualization, Methodology, Software, Investigation, Writing -- Original Draft, Visualization. {\bfseries Mih\'aly Nov\'ak}: Conceptualization, Methodology, Software, Writing -- Review \& Editing, Supervision. {\bfseries Vassil Vassilev}: Conceptualization, Writing -- Review \& Editing, Supervision, Project Administration. {\bfseries Nicolas R. Gauger}: Conceptualization,  Resources, Writing -- Review \& Editing, Supervision, Funding Acquisition. {\bfseries Lukas Heinrich}: Conceptualization, Writing -- Review \& Editing. {\bfseries Michael Kagan}: Conceptualization, Writing -- Review \& Editing. {\bfseries David Lange}: Conceptualization, Writing -- Review \& Editing.

\section{Acknowledgments}
The authors would like to thank members of the MODE collaboration, in particular Tommaso Dorigo, for valuable discussions that helped pave the way to this research project.

MA and NG gratefully acknowledge the funding of the research training group SIVERT by the German federal state of Rhineland-Palatinate. Also, MA and NG gratefully acknowledge the funding of the German National High Performance Computing (NHR) association for the Center NHR South-West.
MK is supported by the US Department of Energy (DOE) under grant DE-AC02-76SF00515. DL and VV are supported by the National Science Foundation under Grant OAC-2311471. LH is supported by the Excellence Cluster ORIGINS, which is funded by the Deutsche Forschungsgemeinschaft (DFG, German Research Foundation) under Germany's Excellence Strategy - EXC-2094-390783311.

This research was supported by the Munich Institute for Astro-, Particle and BioPhysics (MIAPbP), which is funded by the Deutsche Forschungsgemeinschaft (DFG, German Research Foundation) under Germany's Excellence Strategy -- EXC-2094-390783311.
Computing resources have been provided by the Alliance for High Performance Computing in Rhineland-Palatinate (AHRP) via the Elwetritsch cluster at the University of Kaiserslautern-Landau.

\bibliographystyle{ACM-Reference-Format}
\bibliography{main}

\end{document}